\documentclass[12pt]{article}
\usepackage{graphicx}
\begin{document}

\begin{center}

{\large G\"ORAN LINDBLAD IN MEMORIAM}\footnote{For a memorial volume of {\it Open 
Systems and Information Dynamics}.}

\vspace{10mm}

{\large Ingemar Bengtsson}

\vspace{7mm}

{\sl Stockholms Universitet, AlbaNova\\
Fysikum\\
S-106 91 Stockholm, Sverige}

\vspace{5mm}

{\bf Abstract:}

\vspace{5mm}

{\small This is a brief account of the life and work of G\"oran Lindblad.}

\end{center}

\vspace{5mm}

\noindent When quantum mechanics was discovered, in the 1920s, classical 
probability theory was still in a primitive stage. Kolmogorov laid its proper 
foundations in the thirties. Shannon introduced information theory in the forties. 
The fifties and sixties saw developments in the theory of stochastic processes as 
well as in the theory of operator algebras, the works by Stinespring and 
Umegaki being especially relevant to our story. Around 1970 it occurred 
to a small number of physicists, scattered all over the world, that the time 
had come to update quantum mechanics. 

Enter G\"oran Lindblad. He was born in 1940, in the north of Sweden, but his family moved 
south after the war and in 1964 he graduated as an engineer from the Royal 
Institute of Technology in Stockholm.\footnote{Abbreviated as KTH, for Kungliga 
Tekniska H\"ogskolan. G\"oran was always a loyal member of its group for Mathematical 
Physics.} There he found 
Bengt Nagel, a young professor who was building a group in mathematical 
physics. Bengt became G\"oran's PhD supervisor, and set him to work on the 
representation theory of non-compact groups. They achieved some nice results 
together \cite{Nagel}, but half-way through to his PhD G\"oran decided that non-compact 
groups was a crowded field, where only very hard problems remained. He wanted to work 
in a field where there were as few people as possible. Moreover, having read Brillouin's 
book on {\sl Science and Information Theory}, and listened to inspiring lectures by 
Mosh\'e Flato (who was visiting), he was 
interested in Maxwell's demon and in the quantum measurement problem. Bengt 
knew that his very independent student was technically very strong, and 
had no objection to him switching fields. Moreover Bengt had heard of a suitable 
problem: Let $\rho$ be a density operator describing a quantum state, 
let $P_k$ denote projectors associated to a measurement, and let 
\begin{eqnarray} S(\rho ) = - \mbox{Tr}\rho \ln{\rho} \nonumber \end{eqnarray}

\noindent be the information entropy of the state. In the measurement 
\begin{eqnarray} \rho \rightarrow \rho' = \sum P_k\rho P_k = \sum p_k\rho'_k \ , 
\nonumber \end{eqnarray}

\noindent where $p_k = \mbox{Tr}P_k\rho$ is the probability that a particular 
outcome happens. If so the state collapses to the state $\rho'_k$. It was known that 
$S(\rho') \geq S(\rho)$, that is to say that the measurement 
is a dissipative process. The year before Groenewold had conjectured that 
\begin{eqnarray} S(\rho ) \geq \sum p_kS(\rho'_k) \ . \nonumber \end{eqnarray}

\noindent If entropy is identified with ``missing information'' this says that there 
is an average information gain in the measurement process. Bengt suggested that 
G\"oran should prove this conjecture. After a couple of months he did, and published the 
result in the journal that he always referred to, affectionately, as {\sl Communications} 
\cite{Lindblad1}.

This was an encouraging start, and G\"oran now raised his aim. He wanted to adapt 
the concept of information to quantum theory in general and to the measurement process 
in particular \cite{Lindblad2}. The literature he knew was sparse. He did not know that 
Alexander Holevo 
was already in his chosen field, but he studied Umegaki and Stinespring, and a 
few papers by Ingarden and Kossakowski, by Majernik, and by Davies and Lewis. From 
Stinespring he learned about the notion of {\it completely positive maps}, and (unlike 
Stinespring, who focussed on the $C^*$-algebra part of the story) he spelled out their 
physical interpretation very clearly. The maps we are interested in are linear maps taking 
operators to operators. If the operators are density matrices the maps correspond to 
operations performed on quantum states, and they clearly have to be positive in 
the sense that they take positive operators to positive operators. But there may be another 
quantum system, independent of the quantum system on which the operation is performed. 
What Stinespring had observed, and G\"oran was interpreting, is that in this situation an 
operation on the first system that does not influence the second system directly must obey a 
further restriction in order that states for the total system shall be mapped to states. 
It is this restriction that leads to completely positive maps $\Phi$. (Today we would 
say that merely positive maps fail to take the state of the total system to another 
possible state if the total state of the combined system is entangled, but G\"oran did 
not use this word back in 1974.) In fact one can read out of Stinespring's paper that 
completely positive maps arise through the interaction of the system with an external 
quantum system, and are of the form 
\begin{eqnarray} \Phi (A) = \mbox{Tr}_2U^\dagger (A\otimes B)U \ . \nonumber \end{eqnarray}

\noindent The combined system evolves unitarily, but the reduced dynamics of the system 
we are interested in is given by completely positive maps. 

In this way G\"oran convinced himself that there are good physical reasons to 
focus on completely positive maps when building the quantum analogue of Shannon's 
information theory \cite{Lindblad3}. Kraus, and Holevo, had drawn this 
conclusion already, but this G\"oran did not know. 

G\"oran had an application of this in mind. In classical information theory an important 
role is played by the {\it relative entropy} of two probability distributions, and a quantum 
version had been proposed by Umegaki. The importance of the classical relative entropy 
rests on the fact that it cannot increase under stochastic maps. The proof 
of this monotonicity is rather straightforward, but it uses a strategy that fails 
if one tries to prove monotonicity of the quantum relative entropy under completely 
positive maps. G\"oran had seen that it could nevertheless be proved, provided that the 
quantum entropy obeys a subtle property known as strong subadditivity. It does. This was proved 
by Lieb and Ruskai, and G\"oran's proof of the monotonicity of quantum relative entropy 
followed quickly \cite{Lindblad3}. Sometime later, when visiting Paris, G\"oran was 
pleasurably surprised when Elliott Lieb himself contacted him for discussions. Quantum 
information theory was well on its way. 

Another problem that caught G\"oran's fancy was the question of quantum master equations, 
describing the time evolution of quantum systems that interact with an environment (or 
`reservoir', to use the language of statistical physics that G\"oran preferred). 
He studied Hermann Haken's {\it Handbuch} article on laser theory carefully, but as 
a mathematical physicist he was not satisfied with the derivations he found there. 
In the mathematical physics literature that he found, the problem to be solved was 
that of characterizing the generators of a dynamical semigroup. The semigroup was 
supposed to consist of a 
one-parameter family of positive maps $\Phi_t$ obeying the key property that 
\begin{eqnarray} \Phi_s \cdot \Phi_t = \Phi_{s+t} \ . \nonumber \end{eqnarray}

\noindent Hence the state of a physical system at time $s+t$ can be deduced from 
its state at an intermediate time $t$ by evolving the latter during a time $s$. There is 
no memory of the past influencing the motion. This is known as a {\it Markov assumption}. 
It defines an idealized limit where the reservoir has an internal relaxation which 
is much faster than the evolution of the open system we want to describe. Nevertheless 
it is an important case to consider, and the assumption holds in many physically 
interesting situations. 
If suitable continuity assumptions are added one can now rely on 
a mathematical theorem saying that there will exists a densely defined generator 
of the semi-group, such that the evolution of the density matrix becomes 
\begin{eqnarray} \partial_t\rho = {\cal L}(\rho ) \ . \nonumber \end{eqnarray}

\noindent The problem was to characterize the linear map ${\cal L}$ precisely. It 
was still unsolved when G\"oran defended his PhD thesis \cite{Lindblad4}. This was 
in May 1974, Marcel Gu\'enin was the opponent, and of course G\"oran was awarded the 
highest grade. 

The week after the defence G\"oran returned to his desk to consider the problem of 
the generators of dynamical semi-groups.\footnote{When G\"oran told the story he dated 
this to the week after the dinner celebrating the defence. He was fond of good food, 
and would occasionally reminisce about dinners in Paris decades after 
consuming them.} He realized that for physical reasons one should restrict attention to 
completely positive maps. And this is the remarkable thing. Complete positivity, which 
at first sight looks like a complication, makes the structure so much simpler. To get a 
soluble problem G\"oran also strengthened the continuity assumptions somewhat, 
so that he dealt with bounded operators only. With this, he saw that an ``if and 
only if''-statement about the generators of a dynamical semigroup of completely 
positive maps should follow. After a couple of weeks he was as excited as Archimedes, 
but---he wanted to stress when he told the story---he did not go to quite the lengths 
that Archimedes did. The equation he arrived at was 
\begin{eqnarray} L(\rho ) = \frac{1}{2}\sum([V_j\rho,V_j^\dagger ] + 
[V_j,\rho V_j^\dagger ]) - i[H,\rho ] \ , \nonumber \end{eqnarray}

\noindent where ``$L$'' stands for Liouville. In modern notation 
\begin{eqnarray} {\cal L}(\rho ) =  - i[H,\rho ] + \frac{1}{2}\sum([L_j\rho,L_j^\dagger ] + 
[L_j,\rho L_j^\dagger ]) \ , \nonumber \end{eqnarray}

\noindent where the $L_i$ are the Lindblad operators. 

\begin{figure}[t]
\center{
\includegraphics[angle=0,width=0.60\columnwidth]{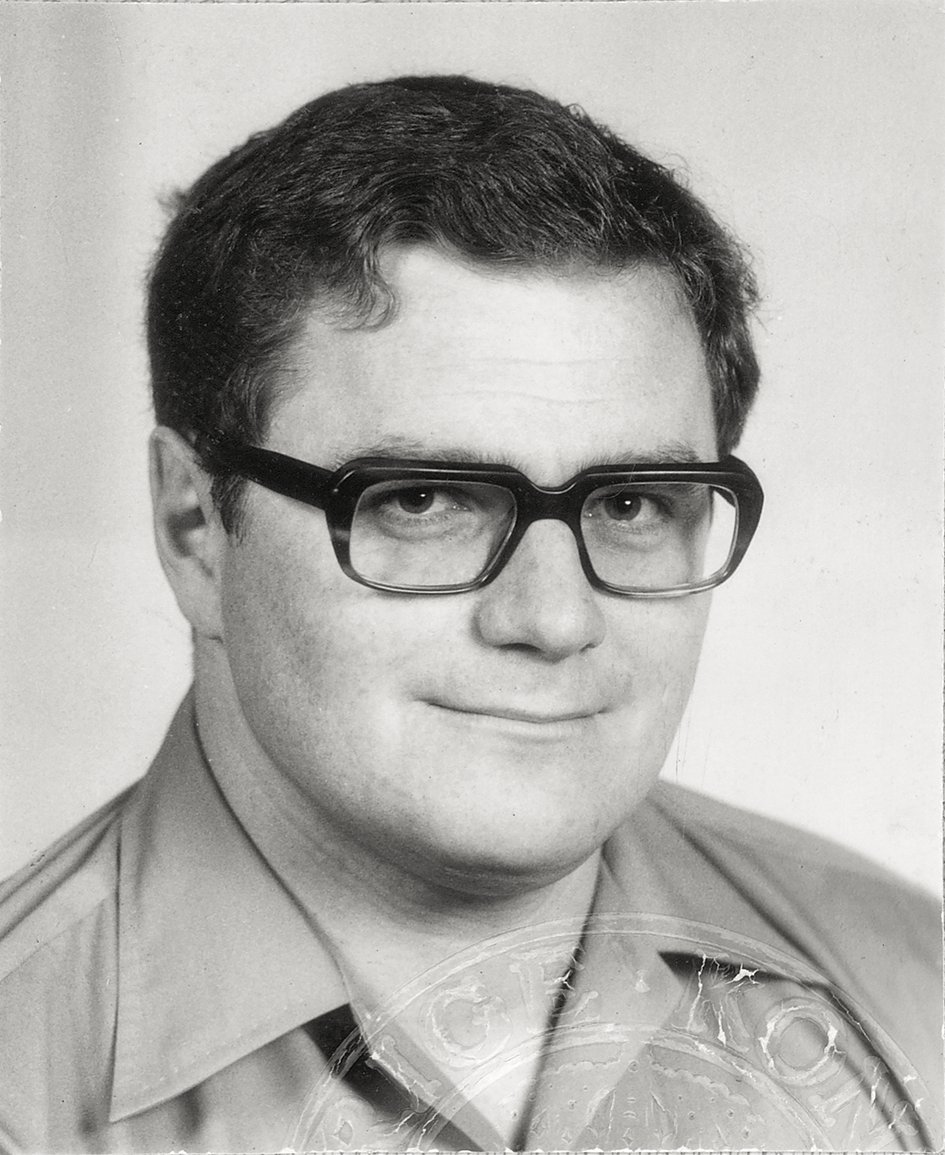}}
\caption{Lindblad in 1974.}
	\label{fig:3}
\end{figure}

G\"oran decided to make his discovery public at a meeting in Toru\'n to which Roman Ingarden 
had invited him. After his talk Ingarden informed him that similar results had been 
derived in Texas by George Sudarshan and his two young visitors, Vittorio Gorini and 
Andrzej Kossakowski. In January the following year Gorini passed through Stockholm on his 
way home, and when comparing their results they found them to be the same up to rearrangement 
of terms, although Gorini and his collaborators had restricted themselves to the finite 
dimensional case. G\"oran's result was more general, and presumably this 
is why the equation became known as the Lindblad equation. 

The equation is now a working horse in many fields of physics, and heuristic derivations 
have even entered undergraduate textbooks. As G\"oran says in his paper, it is an empirical 
fact that the Markovian master equations found in the literature turn out to be of this form 
after some rearrangement of the terms. The point of G\"oran's paper was to show rigourously 
that a bounded generator of a semigroup defined using completely positive dynamical 
maps has to have 
this precise form. It is a rare instance of a result in mathematical 
physics influencing the development of applied areas of physics. In fact, as G\"oran was 
to notice with some contentment, his paper is the most cited paper ever to be published in 
{\sl Communications} \cite{ekvation}. 

After Toru\'n G\"oran travelled widely. Also, Alexander Holevo came to KTH for an extended 
visit, and they became good friends.\footnote{On his bookshelf at home, G\"oran kept an 
Easter egg that was a gift from Holevo.} G\"oran participated in the `Quantum Probability' 
series of conferences organized by Luigi Accardi, and contributed results on quantum ergodicity 
and chaos \cite{Lindblad7}, how to define a non-commutative analogue of the dynamical 
Kolmogorov--Sinai entropy \cite{Lindblad8}, and more. One of his most well known 
results, that of ``entropy exchange'' (which was rediscovered by Schumacher when the time 
was ripe for it), was published in the proceedings of a conference \cite{Lindblad9}. 
However, at some point G\"oran felt that he had heard too many talks by older 
scientists who were just repeating themselves, and he did not want to become one of them. 
From that point on he stayed at home, and declined all invitations to conferences. 

G\"oran made his first serious attack on non-Markovian quantum evolution when 
visiting Bedford College \cite{Lindblad6}. 
Being in England he benefitted from discussions with Brian Davies, and indeed 
his proposal was more influenced by the operational approach to quantum theory than by 
the classical theory of stochastic processes. In the Festschrift for Lamek Hulth\'en 
G\"oran even drew a line all the way back to Bohr, tracing the origin of the stochastic and 
irreversible aspects of the `reduced' quantum dynamics to the fact that we can only 
study one of several possible aspects of the observed system with a single experimental 
setup \cite{Lamek}. Non-Markovian quantum evolution poses difficult problems, and 
G\"oran was to return to them many times over the years \cite{Lindbladsub}. 

He had a lasting interest in statistical physics. Already as a graduate student he 
made an incisive analysis of Maxwell's demon \cite{Lindblad10}, and the demon reappears in the 
one book that he wrote \cite{Lindbladbok}. There the aim is to define the entropy of a system 
out of equilibrium, and to explain where irreversibility comes from. Energy and work now become 
the central concepts. Out of equilibrium the entropy functions are no longer information 
entropies, and are instead defined by the thermodynamic processes allowed by the 
experimenter's control of the system. Here he was influenced by the spin echo experiments, 
and also by many discussions, in Geneva and elsewhere, with Bogdan Mielnik. 
Since his book is discussed elsewhere in this volume we do not go 
into the details here, and instead offer a quote that captures something 
of G\"oran's views on thermodynamics \cite{Lindblad12}: 

\

{\small The irreversibility associated with the second law is sometimes seen as an 
unfortunate imperfection in the deterministic evolution described by the equations of 
motion of QM. But this is a misleading thought. In a world where we cannot control in 
detail the intial state for a large system, the second law allows a more realistic control. 
A flashlight will provide light by the pressing of a button; we need no control of the 
electrons. The button changes a macroscopic boundary condition, and the second law does 
the rest. Truly, the second law is the foundation of our press-button civilization, and 
instruments in the laboratory also work that way.
}

\

\noindent There is of course much more to say about instruments in the laboratory, 
and about the measurement problem. G\"oran did not believe that decoherence in itself 
is sufficient to select a `classical' domain \cite{Lindblad14}. Instead he sought the 
classical aspects of the macroworld in the existence of structures which are departures 
from complete equilibrium but are metastable over a long time scale. His mature views 
on the measurement problem are sketched elsewhere in this volume \cite{Lindblad15}. 

In the 1990s the field of quantum information grew dramatically, but G\"oran no longer 
published much. He made an exception when he saw the ingenuous proof of the no-broadcasting 
theorem, and contributed an approach of his own that exploited the structure of completely 
positive maps in a systematic way \cite{Lindblad16}. It is perhaps also worth mentioning 
that in 2008 he 
published the second paper of his on which he had a co-author \cite{Langmann}. This time it 
was Edwin Langmann, a younger member of the mathematical physics group at KTH. But although 
he did not 
publish much, he worked on many things, including photosynthesis, quantum memory, and 
classical and quantum L\'evy flights. When Matlab became available he developed the 
skill to perform numerical experiments. He attended seminars regularly, occasionally 
intervening on the side of the speaker if he felt that the audience made unreasonable 
demands on mathematical rigour. And whenever an undergraduate student presented a 
project on quantum mechanics G\"oran always found the time to attend, and to make some 
cryptic but encouraging comment. 

His characterisation of quantum mechanics, made in a 
lecture for the undergraduates, is perhaps worth preserving:\footnote{This loses a little 
in translation from the Swedish. Also, it really should be accompanied by the picture of the 
pot that G\"oran drew on the board. He was good at drawing.} 

\

{\small Quantum mechanics is like a pot for making soup. It is almost indestructible, and 
very rigid, but it is also very flexible because you can choose any ingredients you want 
for the soup.
}

\

\noindent His view on the problem of the objectivity of events in quantum mechanics can be 
gleaned from his last paper \cite{Lindblad15}, and perhaps it can also be illuminated 
by another quote which contains a warning for theorists: 

\

{\small We have to recognize that the mathematical formalism of quantum mechanics is not 
a theory of everything. In particular we should not assume that we can classify all 
experiments relevant for this issue; experimental physics is a culture of its own, 
with its own development which is not predictable by theorists.}

\

G\"oran wrote masterful lecture notes for the courses he was teaching, but about the one 
book he published he used to say with a laugh that ``it left no big impression''. 
As is made clear elsewhere in this volume it probably should have, but then, beneath his 
reserved manner, G\"oran had a sunny disposition. He was not worried about such things. He 
was fond of quantum mechanics, and enjoyed thinking about it. Outside physics he was 
deeply interested in history, and read widely in four languages. 

Conversations with G\"oran were always pleasant, and often memorable. To give their flavour, 
I will quote an example that I happen to have in writing. Years ago, when my son was very young, 
G\"oran had told me that one can imagine a superdevice causing laboratory events to 
``unhappen''. I did not quite follow his argument, but I reported it as well as I could at the 
dinner table at home. I then had to write an email to G\"oran, telling him that my son had 
expressed some concern about this possibility. G\"oran's response was: 

\

{\small I do not see any spookiness here. (You can calm your 
son down.) If an event has been observed by an observer it cannot be made to unhappen by 
this observer, only by strictly eventless observers. 

Which reminds me of a black and white jewel, `Grand Hotel' from 1932, with Garbo and others 
(a novel by Vicki Baum). It is full of dramatic events, but in the closing scene a gentleman 
in an armchair in the foyer says: ``Nothing ever happens here!''.}

\

\noindent Perhaps this gives an idea of what it was like to talk to G\"oran. We miss him!

\vspace{15mm}

\noindent \underline{Acknowledgements}: It is a pleasure to thank some of G\"oran's friends for 
reading the manuscript.

\end{document}